\documentclass[twocolumn,prb,showpacs,amsmath,amssymb]{revtex4}
\usepackage{graphicx} 
\usepackage{epsfig}
\newcommand{\br}{\mathbf{r}}

\newcommand{\bn}{\begin{equation}}
\newcommand{\ee}{\end{equation}}
\newcommand{\bga}{\begin{eqnarray}}
\newcommand{\eda}{\end{eqnarray}}
\newcommand{\half}{\frac{1}{2}}
\newcommand{\diff}{\mathrm{d}}

\newcommand{\A}{\mathrm{\AA}}

\newcommand{\bigO}{\mathcal{O}}

\newcommand{\chalmersMC}{Department of Microtechnology and Nanoscience, MC2,
Chalmers University of Technology,
SE-41296 G\"{o}teborg, Sweden}

\begin{document}
\title{Structure and binding in crystals of cage-like molecules:\\ hexamine and platonic hydrocarbons}

\author{Kristian Berland} \affiliation{\chalmersMC}
\author{Per Hyldgaard} \affiliation{\chalmersMC}

\date{\today}

\begin{abstract}
In this paper, we show that first-principle calculations using a van der Waals density functional (vdW-DF), [Phys. Rev. Lett. $\mathbf{92}$, 246401 (2004)] permits determination of molecular crystal structure. We study the crystal structures of hexamine and the platonic hydrocarbons (cubane and dodecahedrane). The calculated lattice parameters and cohesion energy agree well with experiments. Further, we examine the asymptotic accounts of the van der Waals forces by comparing full vdW-DF with asymptotic atom-based pair potentials extracted from vdW-DF. The character of the binding differ in the two cases, with vdW-DF giving a significant enhancement at intermediate and relevant binding separations. We analyze consequences of this result for methods such as DFT-D, and question DFT-D's transferability over the full range of separations. 
\end{abstract}

\maketitle

\section{Introduction}

%\begin{equation}
%E^c_{\mathrm{nl}},\,  E_{\mathrm{vdW}} [\mathrm{meV}]
%\end{equation}

%\bn
%d[\mathrm{\AA}]
%\ee
\begin{figure*}[t]
\centering
\includegraphics[width=15cm]{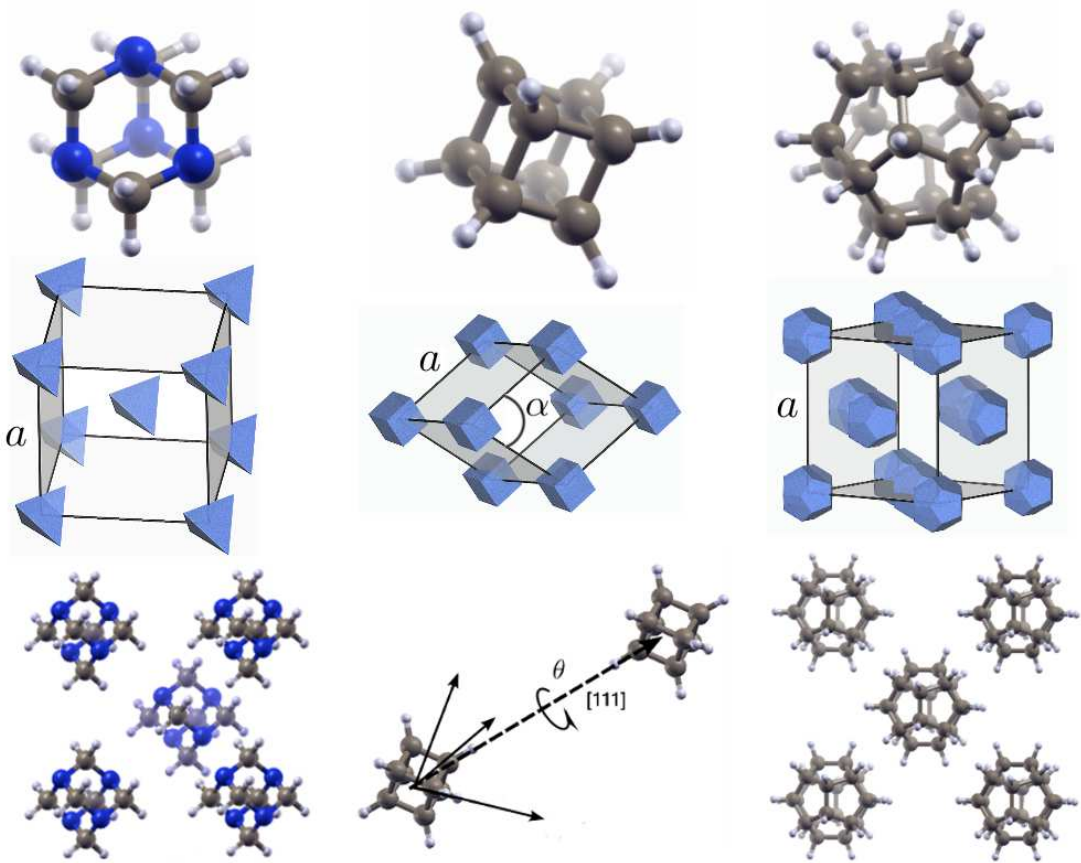} 
\caption{The molecular and crystal structure of hexamine, dodecahedrane, and cubane. The upper panels show the respective molecular structures. The mid panels give schematics of the crystal structures, where the molecular symmetry and orientation are highlighted by the use of a tetrahedron, dodecahedron, and a cube in place of the molecules. The lower panel shows, for hexamine and dodecahedrane, facets of the crystal structure and for cubane the orientation of the cubane molecule in the unit cell. Hexamine has a bcc unit cell, dodecahedron an fcc unit cell, and cubane a rhombohedral unit cell, with equal angles $\alpha$ between the lattice vectors.}
\label{fig:BigMainFig}
\end{figure*}

Understanding supramolecular structure and interactions is essential for understanding many biological processes.\cite{Lehn:Supra} Biological and other supramolecular complexes as polymers and overlayers are sparse matter, that is, they contain low electronic density in essential regions. The general lack of order in these systems prohibits precise measurements of atomic structure and therefore challenges development of theoretical methods. In turn, this makes transferable {\it first principles} schemes attractive; an accurate account of simple periodic structures permits accurate characterization to be made and reliable conclusions to be drawn for more complex (non-periodic) systems. Molecular crystals and simple polymer crystals\cite{Kleis:toward} are ideal testing grounds for applications of first-principle descriptions of sparse matter. Unlike most sparse matter, they constitute ordered systems and therefore leads to unambiguous comparison between theory and experiments.

The dispersion forces underpin the cohesion of sparse matter. Modeling of sparse matter at the electronic level, therefore, requires that we take these effects into account. Traditional implementations of Density-Functional theory (DFT), which parameterizes the density functional within the local density approximation (LDA) \cite{vwn} or the generalized gradient approximation (GGA) \cite{PW91,PBE} have allowed for routine modeling of matter with  dense electron distributions. However, the lack of non-local correlation and hence dispersion forces in these implementations, has effectively inhibited widespread use for general sparse matter systems. Computational methods able to predict molecular crystal structure and stability are also of value in the pharmaceutical industry, for example in mapping out competing crystalline phases for the fabrication of pharmaceuticals.\cite{Phizer} The electronic structure and response are often required to understand and compute properties of technological significance. Consequentially, a range of developments has aimed to extend density functional approximations with an account of dispersive interactions and thus capability to address sparse matter challenges. Such approaches include non-local functionals \cite{other:Dobson,other:Kohn,planar:vdW,Dion:vdW,MIT1,MIT2}, intermolecular perturbation theory \cite{other:Flores1,other:Flores2}, and the addition of a semi-emperical atom-based account of the dispersion forces (DFT-D). \cite{emp:Scoles,emp:Wu,emp:Grimme,emp:Petr} DFT-D has also been used to characterize molecular crystals.\cite{DFT-D-molcrys1,DFT-D-molcrys2}

We study molecular crystal using a van der Waals density functional (vdW-DF).\cite{Dion:vdW} 
The correlation part of this functional depends non-locally on the density and accounts for dispersion forces. 
%ALT: The correlation part of functional accounts for dispersion forces through its non-local dependency on the density.
vdW-DF has been used in a series of first-principle studies of sparse matter, \cite{Review:vdW} yielding for example new insight to the twist of DNA, \cite{vdWDF:DNA} nanotube bundles,\cite{C6:Kleis} water hexamers,\cite{Andre:Water} metal-organic frameworks,\cite{MOF1} and the binding mechanisms at organic-metal interfaces.\cite{vdWDF:Triophene,vdWDF:110,vdWDF:KB,vdWDF:Japan}

In this paper, we first demonstrate that vdW-DF permits structural determination of molecular crystals and second  asses the character of the attractive dispersion interactions. For the first, we calculate the lattice parameters for crystals of the cage-like hexamine, dodecahedrane, and cubane molecules. For the second, we compare an atom-centered asymptotic-$1/r^6$ approximation, frequently encountered in force-field methods \cite{Phizer,molPairRev,CubaneMD} and in DFT-D, 
with the correlation energy provided by vdW-DF. We also analyze consequences for DFT-D and question its performance at intermediate separations. The particular choice of molecules was motivated in part by their aesthetic appeal and in part for the property that their symmetric geometries permits a simple assignment and analysis of the strength of the asymptotic interactions even when expressed in terms of an atom-pair basis.  

This paper has the following plan: The next section presents the molecules hexamine, cubane, and dodecahedrane and their experimental crystal structures. The third section deals with the computational details of vdW-DF. The fourth gives the results for lattice parameters and bulk modulus. In the fifth section, we compare the asymptotic $1/r^6$-form of van der Waals interactions with the full non-local correlation for different molecules, and the vdW-DF potential energy curve with DFT-D calculations for a cubane dimer. The final section holds our conclusions. 

\section{Platonic matter}

Plato asserted that the basic building blocks of Nature were five geometrical structures, today known as platonic solids.\cite{Plato} These are the five convex polyhedra with all faces, edges, and angles congruent. On the molecular level, two of these solids have synthetic hydrocarbon (compounds of only carbon and hydrogen) analogues: cubane (C$_6$H$_6$), which corresponds to the cube, and dodecahedrane (C$_{20}$H$_{20}$), which corresponds to the twelve-faced dodecahedron. The fascination mathematicians held for these geometrical structures since antiquity echoed in the more recent struggle to synthesize their hydrocarbon representatives. The highly strained bonds posed the main challenge. Obstacles were eventually overcome and in 1964 cubane was synthesized\cite{syntCubane} followed by dodecahedrane in 1978.\cite{syntDod} A third, tetrahedrane exists only within a larger chemical structure\cite{Tetra:Pross} and no crystallographic characterization exists. Tetrahedrane does therefore not represent a good testing ground for vdW-DF and we instead study the molecular crystal of hexamine (C$_6$H$_{12}$N$_4$). Although not a hydrocarbon, it shares several features with the platonic hydrocarbons: it is organic, its non-hydrogen atoms form a cage, and this cage has the symmetry of platonic solid tetrahedron. The top row of Fig. \ref{fig:BigMainFig} shows the molecular structures (from left to right) of hexamine, cubane and dodecahedrane, ordered according to the ascendancy of their platonic analogues.  
% with nitrogens forming a tetrahedron with carbon atoms along its edges. 

The second and third row of Fig. \ref{fig:BigMainFig} show the crystal structures. The first column shows hexamine, the mid, cubane, and the final, dodecahedrane. The crystal structure of hexamine forms a bcc with a single molecule in each unit cell and I$\bar{4}3$m symmetry. Hexamine has been used as a model system in numerous studies \cite{Hexamine:ModThermal,Hexamine:ModPressure,Hexamine:ModDielectric} and the crystal structure was determined as early as in 1923.\cite{Hexamine1923}  The crystal structure of cubane is rhombohedral with equal external angles and a single molecule in each cell. The molecule is oriented according to the R$\bar{3}$ space group; a rotation about the [111] axis specifies the configuration of the molecule.\cite{CubaneMD} The crystal structure of dodecahedrane is fcc, also with a single molecule in each cell and Fm$\bar{3}$-space group.\cite{dodExp2} 

Hexamine finds numerous industrial uses.\cite{Hexamine:Industrial} For instance, it serves as a component in fuel tablets and as an antibiotic.\cite{HexEnc} The platonic hydrocarbons find only hypothetical  applications; cubane has been identified as a potential high-energy fuel and explosive.\cite{Cubane:Borman}

In addition to being well-studied crystals, especially hexamine and cubane, these crystals make attractive testing grounds for vdW-DF for two more reasons. First, their simple structures allow for brute-force determination of lattice parameters. Within the crystal symmetry, this determination corresponds to mapping out the potential energy of a single parameter for hexamine and dodecahedrane, and three for cubane. The brute-force approach facilitates post-processing analysis, such as computation of bulk modulus,\cite{bulk:Elini} and makes it easier to evaluate how the choice of exchange in DFT affects crystal structure and cohesion energy. Second, the high symmetry of the molecules reduces the large set of atom-to-atom $C_6$ coefficients to only a few equivalent values. This reduction simplifies the comparison of full vdW-DF calculations with atom-based asymptotic accounts of the van der Waals forces. 

\section{Computational details}

The crystal structure and bulk modulus of hexamine, cubane, and dodecahedrane are determined with DFT using a nonlocal density functional called vdW-DF. The traditional semi-local GGA for exchange-correlation provides accurate bond lengths and charge density $n(\br)$, but fails to capture correlated motion of separated electrons: the long-range dispersion forces. vdW-DF includes these correlations and can therefore account for the structure and cohesion of sparse matter. Since details of the functional and its implementation are given elsewhere, \cite{Dion:vdW, Review:vdW, SelfCons, IJQC:vdW, vdWDF:potassium} we focus mostly on computational steps specific for determination of crystal structures. 

The non-local correlation of vdW-DF takes the form of a double-space integral,
\bn
E_c^{\mathrm{nl}}[n] = \frac{1}{2} \int_{V_0} d\mathbf{r} \int_{V} d\mathbf{r}' 
n(\mathbf{r}) \,
\phi(\mathbf{r},\mathbf{r}') 
n(\mathbf{r}')\,,
\label{eq:Ecnl}
\ee
over an interaction kernel $\phi(\mathbf{r},\mathbf{r}')$. Here $V_0$ denotes the central unit cell and $V$ (formally) the entire space. The kernel can be tabulated in terms of two parameters $d$ and $d'$ related to the local response $q_0(\br)$ and spatial separation $|\br-\br'|$ by  $d=q_0(\br)|\br-\br'|$ and $d'=q_0(\br')|\br-\br'|$. The remaining part of the exchange-correlation functional of vdW-DF consists of the exchange part of revPBE \cite{revPBE} and the correlation of LDA:
\bn
E^{\rm vdWDF}_{\rm xc}=E_c^{\rm LDA}+E_x^{\rm revPBE} + E_c^{\mathrm{nl}}[n]\,.
\ee

The total energy functional of vdW-DF, $E^{\rm vdWDF}[n]$, also includes the standard elecrostatic and kinetic-energy terms within the Kohn-Sham scheme.\cite{KS} It is convenient to write this energy in terms of a semi-local part $E_0[n]$ containing all but the non-local correlation, so that $E^{\rm vdWDF}[n]=E_0[n]+ E_c^{\mathrm{nl}}[n]$. For input charge density $n(\br)$, we use the result of semi-local calculations with the PBE\cite{PBE} flavor of GGA.  We will refer to calculations with the PBE flavor of GGA as DFT-GGA. The charge density could also have been obtained within fully self-consistent vdW-DF.\cite{SelfCons,Soler:speedup} However, the two-step non-self consistent procedure introduces only a slight approximation. Previous studies have documented that for systems with small charge transfer, binding energies of non-self consistent vdW-DF only differ by tiny amounts  from fully self-consistent energies.\cite{SelfCons,benzeneWater}

To speed up evaluation of the non-local correlation, we introduce a radius cutoff based on the decay of van der Waals forces at large separations. With this cutoff, the kernel takes form
\bn
E^{nl}_c[n] \approx \half  \int_{V_0} \diff \br\, n(\br) \int_{|\br-\br'|<R}  \diff \br' \,   \, \phi(\br,\br') n(\br')\,. \label{Enl:vdWDFrad}
\ee
In the above expression, we see that the CPU-cost for evaluating the non-local part of vdW-DF goes as $R^3\bigO(V_0)$. Thus, for large or periodic systems the computational costs increase linearly with system size. To cut computational costs further, we introduce an extra radius cutoff corresponding to the separation between dense and sparse sampling of the charge-density grid. We note that Ref.~\onlinecite{Soler:speedup} reports a more elaborate scheme  which considerably reduces CPU-costs, yet our simple measure was sufficient for our non-self consistent calculations as the underlying DFT-GGA calculations of electronic density dominated time consumption. 

The use of revPBE for exchange in vdW-DF was motivated by the fact that this exchange functional excludes unphysical binding effects at large distances. \cite{IJQC:vdW,exchange:Wu} For a range of systems, vdW-DF overestimates binding separations. \cite{IJQC:vdW,Review:vdW} Several studies indicate that this discrepancy can be attributed to the details of the exchange functional. \cite{Review:vdW,benzeneWater,exchange:Puzder,exchange:Risto,
KlimesComment}  Puzder {\it et al} \cite{exchange:Puzder} have demonstrated that replacing Hartree-Fock exchange with revPBE improves binding separations for benzene dimers. Gulans {\it et al}\cite{exchange:Risto} have shown that for a selected range of molecular complexes the PBE exchange functional improves binding energies. We furthermore illustrate the sensitivity to exchange by including results based on use of an alternative vdW-DF(PBE), where revPBE exchange has been replaced by that of PBE. We do not argue for replacing vdW-DF with vdW-DF(PBE), instead we simply explore consequences of a different account of exchange.\cite{KlimesComment}

To calculate crystal parameters and cohesion energy, we minimize the potential energy. In many respects this vdW-DF structure determination is similar to those in Refs.~\onlinecite{vdWDF:extended,C6:Kleis,Kleis:toward,vdWDF:potassium}. The potential energy is given by the difference between the total energy of the full crystal and a reference energy for a system of isolated molecules
\begin{equation}
E_{\rm coh}(a,\{\alpha\theta\})= E^{\rm vdWDF}(a,\{\alpha\theta\})- E^{\rm vdWDF}(a\rightarrow \infty,\{\alpha\theta\}) \, . 
\end{equation}
In the above equation the curly brackets are specific for the cubane crystal as it depends on three rather than one parameter. In the reference calculation corresponding to the reference energy, $E^{\rm vdWDF}(a\rightarrow \infty,\{\alpha\theta\})$, the semi-local part is obtained somewhat differently from the part containing the non-local correlation. For the former, 
we effectively isolate the molecules in our periodic boundary calculations, by using a unit cell of doubled size in all directions. This measure secures negligible charge overlap between the molecules in the supercells. For the latter (nonlocal) part we restrict the integral of Eq. \ref{Enl:vdWDFrad} to the central supercell to avoid coupling between the enlarged unit cells. Hence, only non-local correlations within the molecule contribute to the reference energy.  

To enhance accuracy in the evaluation of the non-local part of the potential energy, we systematically cancel a small, but noticeable, grid dependence in the evaluation of Eq. \ref{Enl:vdWDFrad}. This cancellation is performed by making sure to use the same FFT grid spacing in the reference calculation as in the main calculation. Furthermore, we make sure to place the isolated molecules in the same relative configuration to the underlying grid in the reference calculation as in the main calculation. We thus perform an additional reference calculation for every molecular configuration investigated. These measures have been used to secure a high accuracy of the non-local part of vdW-DF in several earlier studies. \cite{vdWDF:potassium,vdWDF:extended,IJQC:vdW,Kleis:toward}

We map the potential energy landscape by varying the lattice parameters of the molecular crystals within the experimental crystal symmetry. The stiff cage-molecules allow us to keep the internal coordinates of the molecules frozen for all configurations. The molecular structures are determined in isolation using the PBE flavor of GGA. The resulting bond lengths will be compared with experimental data in the next section to verify the utility of conventional DFT-GGA for the internal structure of strained molecules. 

\begin{table}[h]
\begin{ruledtabular}
\caption{Experimental and calculated bond lengths of hexamine, cubane, and dodecahedrane. The calculations were done with the PBE flavor of GGA. $l$ denotes the C-C bond length for cubane and dodecahedrane and the C-N bond length for hexamine. $l_{CH}$ denotes the carbon-hydrogen bond length.}
\begin{tabular}{lrrr}
Parameter& Hexamine & Cubane &Dodecahedrane \\
\hline
$l$[\AA] & 1.472 &1.566 & 1.549  \\
$l^{\mathrm{exp}}$[\AA] & 1.476\footnotemark[1] &1.562\footnotemark[2]& 1.544\footnotemark[3]\\ 
$l_{\rm CH}$ [\AA] & 1.101 & 1.095  &1.100 \\
$l_{\rm CH}^{\mathrm{exp}}$[\AA]& 1.088\footnotemark[1] &1.097\footnotemark[2] & - \\  
\end{tabular}
\footnotetext[1]{Ref.~\onlinecite{hexamineCrysExp}.} \footnotetext[2]{Ref.~\onlinecite{cubaneExp}.} \footnotetext[3]{Ref~\onlinecite{dodExp1}.}
\label{table:Bond}
\end{ruledtabular}
\end{table}
The electronic-structure calculations rely on the plane-wave code \verb DACAPO \, \cite{DACAPO} using ultra-soft pseudopotentials. In combination with a separate reference calculation as previously discussed, we secure the convergence of the non-local correlation by specifying an FFT-grid spacing less than 0.13 $\A$. This spacing leads to an effective plane-wave energy cutoff of at least 500 eV. For all crystals, we set the Monkhorst-Pack $\mathbf{k}$-sampling to $4\times 4\times 4$.

\section{Structure determination}
\begin{table}[t]
\begin{ruledtabular}
\caption{The vdW-DF prediction of lattice parameters, cohesion energy and bulk modulus for the crystals of hexamine, cubane and dodecahedrane compared with an alternative vdW-DF(PBE) based on PBE-exchange and with experimental values. The experimental lattice parameters are based on low temperature measurements, except for dodecahedrane.}
\begin{tabular}{llrrrr}
 &Parameter &    vdW-DF  & vdW-DF
(PBE)   & Exp. \\
\hline
{\it Hex.}\\ %&&  &   &  \\
&$a$ (\AA)  & 7.14  &  6.93  &   6.910\footnotemark[1]   \\
&$E_{coh}$(eV) & -1.0127 & -1.427 & -0.827\footnotemark[2]  \\ 
&$B_0 (\mathrm{GPa})$ & 10.0  & 14.0  &  7.0\footnotemark[3] \\   % This number is quite low
\hline
{\it Cub.}& &  &   &  \\
&$a$ (\AA)  & 5.45  &  5.25  &   5.20\footnotemark[4]   \\
&$\alpha$  & 73 & 72.5  & 72.7\footnotemark[4]\\
&$\theta$  & 47.5 & 46.5 & 46 \footnotemark[4]\\
&$E_{coh}$(eV) & -0.77 & -1.15 & -0.857\footnotemark[5]  \\
&$B_0(\mathrm{GPa})$ &  7.2  & 14.8 &    - \\
\hline
{\it Dod.}& &  &   &  \\
&$a$ (\AA)  & 10.92  &  10.56  &  10.60\footnotemark[6]   \\
&$E_{coh}(eV)$ & -1.46 &-2.06 &   \\
&$B_0 $(GPa) & 12.2  & 18.6  &  -  \\
\end{tabular}
\footnotesize{
\footnotemark[1]{Ref.~\onlinecite{hexamineCrysExp}}.
\footnotemark[2]{Ref.~\onlinecite{Hexamine:Cohesion}}.
\footnotemark[3]{Ref.~\onlinecite{Hexamine:Bulk}}.
\footnotemark[4]{Ref.~\onlinecite{hexamineCrysExp}}.
\footnotemark[5]{Ref.~\onlinecite{ChemRef}}.
\footnotemark[6]{Ref.~\onlinecite{Leo:Dod}}}.
\label{table:Crystal}
\end{ruledtabular}
\end{table}

\begin{figure*}[t]
\centering
\includegraphics[height=6.2cm]{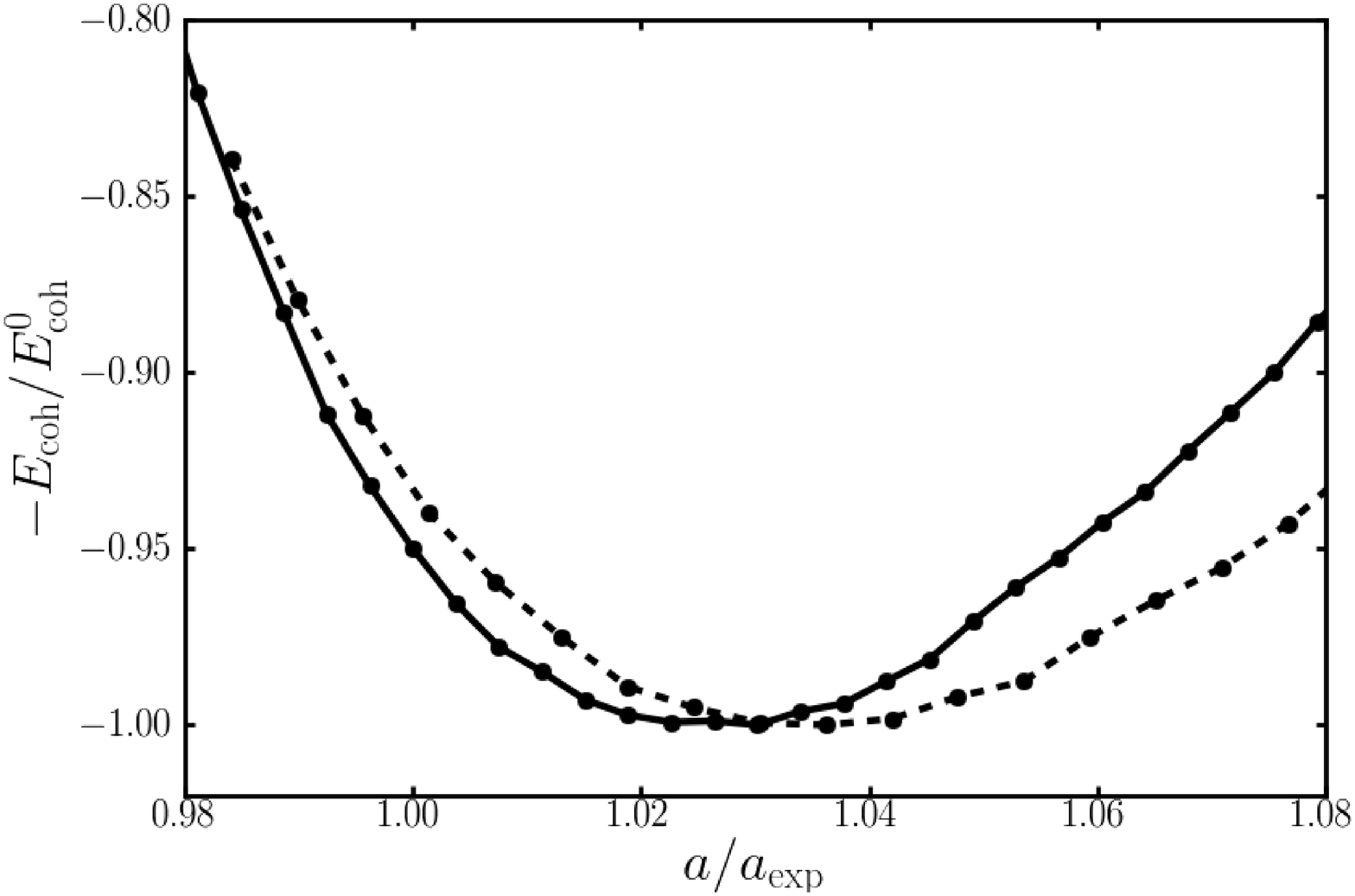} \includegraphics[height=6.2cm]{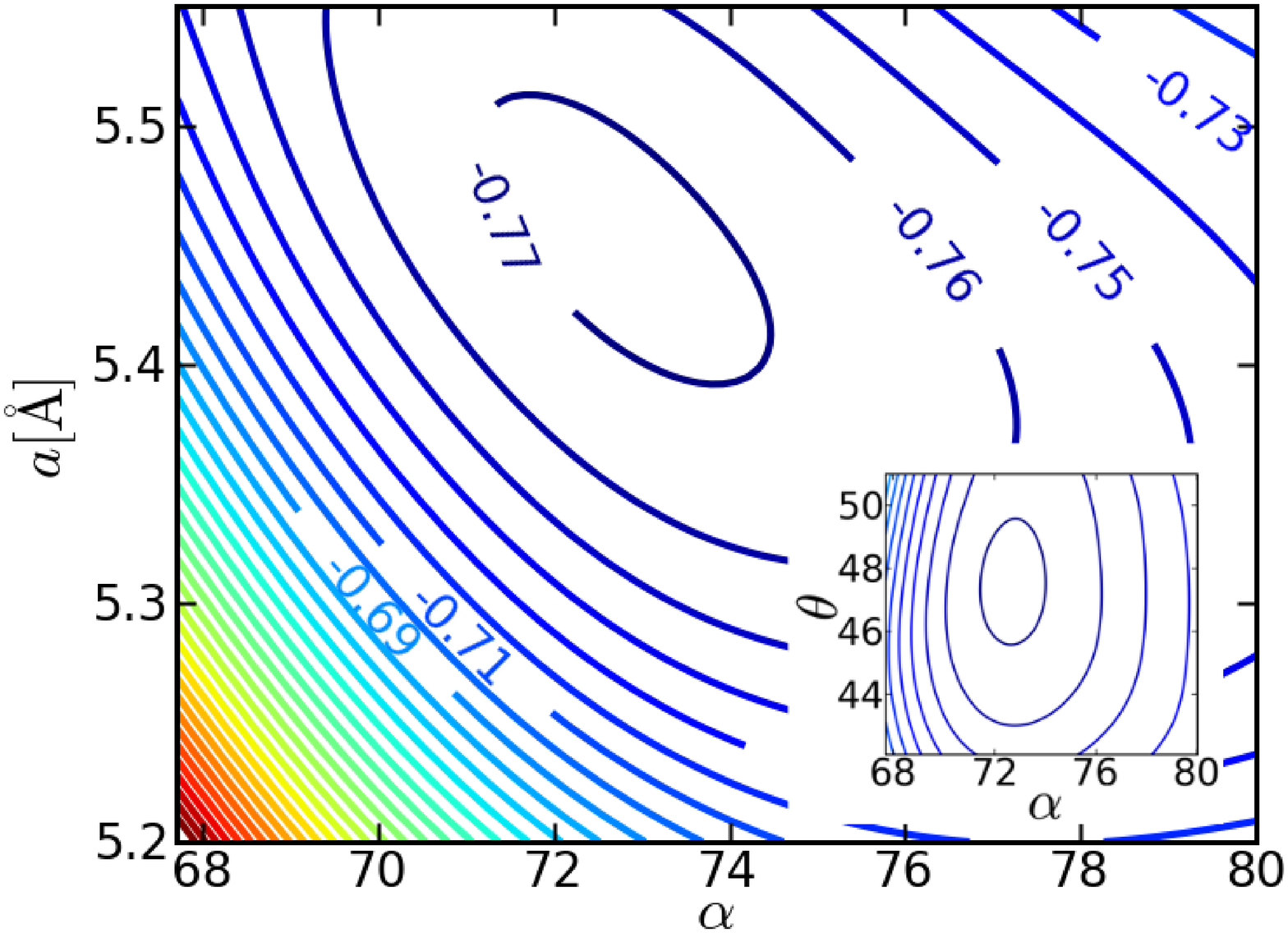}
\caption{The potential energy curves for the crystal of hexamine and dodecahedrane (left panel) and corresponding contour plots for the crystal of cubane (right panel). The left panel shows 
 potential energy where the curves are normalized separately so that the experimental lattice parameter and the calculated cohesion energy equals unity. The solid curve represents dodecahedrane and the dashed hexamine. The difference in the optimal $a/a_{\rm exp}$ value can be attributed to the experimental lattice parameter being measured at low temperature for hexamine, but not for dodecahedrane. The right panel shows spline-interpolated contour plots of the two-dimensional intersection of the three-dimensional potential energy landscape. The main figure corresponds to the optimal value of the internal angle $\theta$, while the insert corresponds to the optimal value of the unit cell length $a$. The pronounced asymmetry of the curves and contours reflect the hard wall provided by Pauli repulsion.}
\label{fig:Binding}
\end{figure*}

\subsection{Molecular structure}

Table \ref{table:Bond} shows the calculated molecular structures. It demonstrates DFT-GGA can account for the intramolecular bonding even in the highly strained cubane molecules. The resulting bond lengths differ by less than 1 \% from the experimental values.  In first principle studies of molecular crystals, accurate determination of lattice parameters require accurate account of constituent molecules. 

Unlike many calculations, where determination of crystal structure often starts from the structure of the individual molecules, most experiments resolve the molecular structure by looking at the diffraction pattern of a full molecular crystal. For our purposes of testing the first-principle vdW-DF method, this is fortunate, as efforts to characterize molecular structures also generate an abundance of experimental data on molecular crystal structures.

\subsection{Crystal structure}

Figure \ref{fig:Binding}. shows the binding curves and contours found by varying the molecular crystals' lattice parameters identified in Fig. \ref{fig:BigMainFig}. The simple crystal structures of hexamine and dodecahedrane give a one-dimensional potential energy landscape, while the cubane crystal has a three dimensional one. In the right panel of Fig. \ref{fig:Binding}, we display the $\alpha a$ and the $\alpha \theta$ intersections, for the optimal values of $\theta$ and $a$. The curves exhibit a pronounced asymmetry around their minimum. This asymmetry arise because the van-der Waals attraction is much softer than the kinetic-energy repulsion.

Table. \ref{table:Crystal} contains the calculated results and experiment values. Standard vdW-DF performs well both for lattice parameters and cohesion energy. If not directly available, we obtain experimental cohesion energies by correcting for gas phase and vibrational contributions to the enthalpy of sublimation, using the method described in Refs. \onlinecite{vdWDF:Nabok,method:cohesive}.  

The bulk modulus is obtained with use of polynomial interpolation according to the scheme of Ziambaras and Schr\"oder.\cite{bulk:Elini} The required polynomial were constructed using data from selected one-dimensional deformations.\cite{bulk:CRYSTAL} For hexamine, where the experimental bulk modulus is available, the computed value show fair agreement with the experimental value. There is also a trend for bigger cage molecules to have a larger bulk modulus. We attribute this trend to the fact that for bigger molecules a smaller relative part of the unit cell consists of soft intra-molecular regions. Therefore, as the relative unit cell dimensions change, the distance between (stiff) molecules changes more for big molecules than for small molecules. 

For all three crystals, we find unit-cell volumes somewhat larger than the experimental ones. Similar overestimations have also been encountered in previous studies.\cite{Review:vdW,IJQC:vdW,KlimesComment} The alternative choice of PBE as exchange functional, vdW-DF(PBE), influences results substantially. On one hand, it improves lattice parameters, almost to level of standard DFT for intra-molecular bonds. On the other hand, the value of cohesive energy and bulk modulus worsens. These results signal that the main discrepancy between experiments and vdW-DF stems from the specific form of semi-local exchange.\cite{Review:vdW,exchange:Puzder,exchange:Risto,KlimesComment} 

We note that an {\it ab initio} study of cubane has previously been performed at the LDA level.\cite{DFTCubane} Being a local functional, LDA has no physical basis for the van der Waals binding that provides the cohesion of this molecular crystal. The spurious LDA binding arises from an unphysical accounts of exchange.\cite{IJQC:vdW,exchange:Wu} Once the molecular crystals are investigated with DFT-GGA, which has an improved account of exchange, the binding essentially vanishes.\cite{DFTCubane} 

\section{Asymptotic pair potentials versus non-local correlation}

\begin{figure}[h]
\centering
\includegraphics[width=9cm]{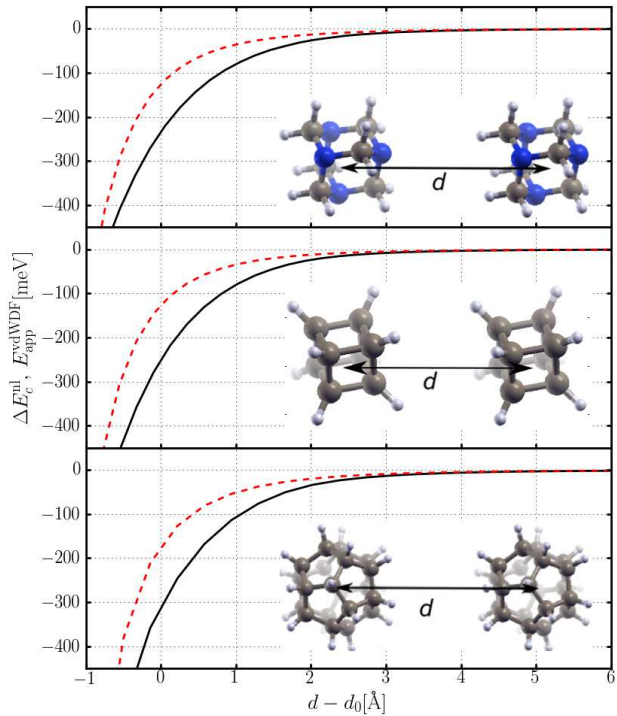}
\caption{Comparison between the non-local correlation of vdW-DF and atomic pairs potentials generated with asymptotic vdW-DF for a dimer of hexamine, dodecahedrane, and cubane (from top to bottom) in configurations corresponding to nearest neighbors in their respective crystal. The dashed curve gives the APP-vdWDF result, while the solid curve gives non-local correlation of vdW-DF: $\Delta E_{\rm c}^{\rm nl}(d)=E_{\rm c}^{\rm nl}(d)- E_{\rm c}^{\rm nl}(d\rightarrow \infty)$.  The horizontal axis gives the difference from the vdW-DF crystal binding separation $d_0$, which respectively takes the values 6.18, 5.45 and 7.72  $\A$.
The difference between the two curves demonstrates the enhancement of non-local correlations over the vdW-DF asymptotic atom-based account at relevant binding separations and intermediate separations. }
\label{fig:C6}
\end{figure}

The asymptotic van der Waals interactions between two atoms or molecules goes as the simple power law $C_6/r^6$, where $C_6$ gives the strength of the interaction. This familiar result can be derived from second-order perturbation theory,\cite{C6:Standard} or from 
an analysis of the shifts in the zero-point motions of the electron.\cite{C6:Mahan} A common strategy in force-field methods and empirical extensions of DFT is to adopt such an asymptotic form at all separations in terms of atom-centered pair-potentials (APP). However, we can not take for granted that the asymptotic behavior should hold for separations closer to that of intramolecular binding. On the contrary, because  van der Waals forces arise from correlated motion of electrons and not from the atomic nuclei, several mechanisms affect and enhance the interaction at short and intermediate separations: higher order moments contribute, polarizability changes as charge is distorted and finite-$\mathbf{k}$ dispersion of the electronic response becomes important. Zaremba and Kohn\cite{ZarKohn} considered adsorption of noble atoms on surfaces and documented a significant enhancement of dispersion energies over an atom-centered account; their asymptotic $1/d^3$ form use the distance to an image plane $d$, rather than to the surface atoms. Within vdW-DF, Kleis {\it et al} \cite{C6:Kleis} demonstrated that for interactions in nanotube bundles, the force  stems primarily from the electron tail around the nanotube, and that as the tubes get closer, higher order moments dominate over the asymptotic interaction. Part of this enhancement can be interpreted as an image-plane effect. 

Here we investigate whether an atom-centered $1/r^6$ form is a good approximation for the non-local correlation of molecular dimers. As our argument is based on the asymptotic vdW-DF account of van der Waals forces, we also discuss for these molecules the accuracy of the asymptotic account. In the first subsection, we compare the non-local correlation of vdW-DF for dimers of hexamine, cubane, and dodecahedrane with its corresponding atom-centered pair potentials. We document a significant enhancement of the non-local correlation at short (binding) separations and at intermediate separations, one~to~three~\AA ngstr\"om \,beyond typical binding separations. In the second subsection, we discuss the accuracy of our $C_6$ coefficients. In the third, we investigate consequences of this result, in particular for the use of DFT-D. We will argue that although standard DFT-D methods can provide good descriptions of both short and asymptotic separations, their asymptotic atom-centered form does not describe the enhancement of correlation at intermediate separations exhibited by vdW-DF. 

\subsection{Asymptotic vdW-DF}

For the the asymptotic van der Waals forces, the  $C_6$ coefficient between two fragments, $A$ and $B$, can be computed from the general formula
\bn
C_6^{AB}= \frac{3}{\pi}\int_0^\infty \diff u \, \alpha_A(iu) \alpha_B(iu) \,, \label{eq:C6}
\ee
where $\alpha(\omega)$ is the polarizability of the fragment. In order to calculate the C$_6$ coefficients we approximate $\alpha(\omega)$ with the local external-field susceptibility of vdW-DF, 
\bn
\chi^{\rm vdW-DF}_A(\omega, \br) = \frac{n_A(\br)}{\left[9q_0(\br)^2/8\pi \right]^2 -\omega^2}\,, \label{eq:response}
\ee
and a polarizability given by 
\bn 
\alpha^{\rm vdW-DF}_A(\omega)= \int \diff^3 \br \,\chi^{\rm vdW-DF}_A(\omega, \br)\,.
\ee

To generate vdW-DF based atom-centered pair potentials (APP-vdWDF), we first partition the full charge density $n(\br)=\sum_i n_i(\br)$ among the atoms of the molecules with aid of Bader analysis. \cite{Bader:main,Bader:alg} Based on this charge partition, we calculate the atom-to-atom $C_6$ coefficients using Eq. (\ref{eq:C6}). Initially, this procedure generates $N^2$ different $C_6$ coefficients for a molecular dimer of $N$ atoms per molecule. For a dodecahedrane dimer, we get as much as $40\times 40 =1600$ coefficients. Fortunately, because of the high symmetry of the isolated molecules, this number reduces to only three equivalent values for cubane and dodecahedrane: $C_6^{\rm C-C}$, $C_6^{\rm C-H}$, and $C_6^{\rm H-H}$. For hexamine, the extra nitrogen atoms lead to six coefficients. As noise in the electronic density affects the value of the coefficients, we average over a large set of equivalent values to obtain the final values.
\begin{table}[h]
\begin{ruledtabular}
\caption{
Computed values of the $C_6$ coefficients (Hartree atomic units) for different pairs of atoms within respective molecules, using the asympotic form of vdW-DF with charge density as partitioned with a Bader analysis. The figure also shows the molecule-molecule $C_6$  coefficients obtained with the Andersson-Langreth-Lundqvist (ALL) scheme,\cite{C6:Ylva} and the molecule-molecule $C_6$ coefficents obtained with the use of parameters given in DFT-D schemes. \cite{emp:Wu,emp:Grimme,emp:Petr}
}
\begin{tabular}{lrrr}
$C_6^{\mathrm{vdWDF}} $ & hex & cub &dod  \\
\hline
C-C  & 4.44  & 13.5 & 11.2  \\
C-H  & 4.04  & 6.87 &  6.45 \\  
H-H  & 3.98  & 3.72 & 4.07  \\  
N-N  &32.6 & -   &  -       \\
N-C  &12.0 & -   &  -       \\
N-H  &11.0 & -   &  -      \\
mol-mol (vdW-DF) &3470 & 1990& 11300  \\
\hline
mol-mol (ALL)    &3270& 1940  & 10200   \\
mol-mol (Wu)    & 4340 & 2600 & 16300 \\
mol-mol (Grimme)& 4000 & 2630 & 16400  \\
mol-mol (Jure\v{c}ka) & 4790 & 3120 & 19500
\end{tabular}
\label{table:C6}
\end{ruledtabular}
\end{table}

The upper part of Table \ref{table:C6} shows the $C_6$ coefficients calculated within asymptotic vdW-DF, both as partitioned according to the Bader analysis and as evaluated for the entire molecule. 
The calculated coefficients per atomic pair deviates much from a naive assignment of the full coefficient of the molecule according to the number of valence electrons of the underlying atom. In such a scheme the C-C coefficient would be 16 times larger than the H-H coefficient, while in fact, for cubane and dodecahedrane, it is merely three-four times stronger. This result can be attributed to the relative stronger response of the low-density regions surrounding the hydrogen atoms as $q_0(\br) \propto [n(\br)]^{1/3}$ (in the homogeneous limit), and these areas dominate $E^{\rm nl}_{\rm c}$.  The somewhat anomalous values for hexamine can be attributed to our Bader analysis scheme partitioning a significant portion of the charge density near the carbon atoms to the centrally located nitrogen atoms.\cite{comment3} For cubane and dodecahedrane the partitioning was similar, and the differences in atomistic $C_6$  values show that they are influenced by their local enviroment. 

Having generating C$_6$ coefficients appropriate for a comparison between the asymptotic account and the full correlation of vdW-DF, we study dimers of hexamine, cubane and dodecahedrane at different separations. We choose orientations that are given by the nearest-neighbor configurations in the crystals. The total asymptotic non-local correlation of APP-vdWDF reads
\begin{equation}
E^{\rm vdWDF}_{\rm app} = \sum_{i} \sum_j \frac{C_6^{ij}}{|\br_i-\br'_j|}\,,
\end{equation}
where $i$ and $j$ label atoms in separate molecules of the dimer. 

Figure \ref{fig:C6} shows results for the full non-local correlation of vdW-DF (solid curve) and that of APP-vdWDF (dashed curve). At large separations these two curves converge. In contrast, they differ significantly at relevant binding (short) separations and at intermediate separations. At these separations the non-local correlation is almost double as large as that of APP-vdWDF, for both hexamine and cubane, while for the biggest molecule, dodecahedrane, the difference is smaller. For all dimers, we also need to go to relatively large separations to recover asymptotic values for the non-local correlation. 

As vdW-DF is an {\it ab initio} functional, based on a set of exact sum rules,\cite{Dion:vdW} the strong enhancement of non-local correlations at short and intermediate separations indicates that the asymptotic form neglects important contributions. %This results could have implications for the construction of atom-based accounts of the non-local correlation.
It also highlights that in constructing modified semi-empirical van der Waals functionals,\cite{MIT1,MIT2} a fitting of the asymptotic functional to $C_6$ coefficients does not guarantee an accurate description at short binding separations. 

\subsection{Comparison of $C_6$ coefficients}

The lower part of of Table \ref{table:C6}, gives the molecular $C_6$ coefficients calculated with asymptotic vdW-DF, the Andersson-Langreth-Lundqvist (ALL) scheme,\cite{C6:Ylva} and computed using the atomic coefficients of Wu~{\it et al}, \cite{emp:Wu}  Grimme,\cite{emp:Grimme} and Jure\v{c}ka~{\it et al}.\cite{emp:Petr}
The ALL and vdW-DF give coefficients smaller than that used in DFT-D schemes. 

We expect that Grimme and Wu provide good values for molecular $C_6$ coefficients, because their underlying atomistic coefficients were fitted to reproduce a range of accurate molecular $C_6$ coefficients calculated from experimental molecular polarizabilities (Ref. \onlinecite{emp:Wu} and references therein). The coefficients of Jure\v{c}ka,\cite{emp:Petr} give somewhat larger molecular $C_6$ coefficients. This comes from the use of Slater-Kirkwood average\cite{SK:Average} for $C_6$ coefficients between different atomic species, while keeping those of Grimme for indentical atomi species (the $C_6$ coefficients of Grimme are optimized for a different average. %The modification of Jure\v{c}ka a  therefore an unintentionally scale the attractive dispersion curve. However, we believe this issue can be mitigated, and for simi to describe both the asymptotic and binding regions. 

Asymptotic vdW-DF and ALL likely underestimates the $C_6$ coefficients for these molecules; they are  20-30 \% smaller than that used in DFT-D methods. For the similar ALL scheme,  Ref.~\onlinecite{C6:YlvaRydberg} reports an underestimation of $C_6$ coefficients for larger molecules, in particular for benzene and C60. A difference between the ALL scheme and asymptotic vdW-DF is that for the former a hard cutoff accounts for plasmon damping, while for the latter, the local response $q_0(\br)$ provides a smooth cutoff. There is good consistency between the two methods. Both methods also assume a local, scalar, relationship between the applied and the full electric field, which is an approximation for finite-sized objects.\cite{C6:Ylva} We speculate that this approximation contributes to the underestimation of $C_6$ coefficient for the investigated, relatively large, molecules.

%Ref.~\onlinecite{C6:YlvaRydberg} documents that the ALL scheme underestimates the $C_6$ coefficients of large molecules. Asymptotic vdW-DF and ALL differ in that 
  %Both vdW-DF and ALL underestimate molecular $C_6$ coefficients by 20-30 \% compared to that computed using the atomistic $C_6$ coefficients used in DFT-D methods. In the next part of this section, we discuss these results. 

\begin{table}[h]
\begin{ruledtabular}
\caption{Atomistic $C_6$ coefficients for cubane used in APP-vdWDF (calculated with asymptotic vdW-DF and charge density partitioned with Bader analysis), and coefficients used in DFT-D methods.}
\begin{tabular}{lrrrr}
$C_6^{\mathrm{vdWDF}} $ & vdW-DF & Wu\footnotemark[1] &Grimme\footnotemark[2] &  Jure\v{c}ka\footnotemark[3] \\
\hline
C-C  &  13.5&  22.06   &28.3    & 28.3  \\
C-H  &  6.8 &  7.89    & 5.01  & 8.82  \\  
H-H  &  3.7 &   2.83   & 2.75  & 2.75  \\  
%mol-mol  &1990&  2600  & 2630  & 3120  \\
\end{tabular}
\footnotetext[1]{Ref.~\onlinecite{emp:Wu}}.
\footnotetext[2]{Ref.~\onlinecite{emp:Grimme}}.
\footnotetext[3]{Ref.~\onlinecite{emp:Petr}}.
\label{table:C6cubane}
\end{ruledtabular}
\end{table}

Table \ref{table:C6cubane} shows atomistic $C_6$ coefficients for cubane as calculated with asymptotic vdW-DF, and given by Wu,\cite{emp:Wu} Grimme,\cite{emp:Grimme} and Jure\v{c}ka,\cite{emp:Petr} for use in DFT-D schemes. vdW-DF weights the relative response of the carbon less than that of the hydrogen, compared with the values of DFT-D.\cite{comment3} This property could relate to the above mentioned approximate treatment of electrodynamics. It could also relate to the carbon atoms being located somewhat inside the molecule, having a different local charge density and responding less to external fields than an atom on the exterior would; in contrast, DFT-D does not discriminate between atoms at different locations.

%To check whether the enhancement at shorter separations is an artifact of the specific atomistic assignment, we plot, in the top panel of Fig. \ref{fig:C62}, the APP curve (thick dashed line) against a modified APP (thin dashed line). The modified APP (modAPP) divides the full asymptotic vdW-DF interaction among the atoms, using the same relative strength as Grimme. That the modified APP curve gives a weaker attractions than APP, shows that the relative assignment of the interaction strength does not cause the enhancement. 
%

\subsection{Consequences for atom-based pair potentials: the binding curve of cubane}

The vdW-DF results presented in the first subsection shows that a simple asymptotic account only partially captures correlation effects at short and intermediate separations. As DFT-D use such an asymptotic form to describe non-local correlations, this result stands in apparent contrast to the many successful applications of DFT-D.\cite{emp:Wu,emp:Grimme,emp:Petr,DFT-D1,DFT-D2,DFT-D3} 

To understand consequences of our result for methods such as DFT-D, we must first consider other effects that could contribute to the difference between APP-vdWDF and the non-local correlation of vdW-DF.  In vdW-DF, correlations are described by $E_{\rm c}^{\rm LDA} +E_{\rm c}^{\rm nl}$ and hence $E_{\rm c}^{\rm nl}$ also accounts for semi-local correlations.\cite{Dion:vdW,IJQC:vdW}  Second, we must consider the specific designs of actual DFT-D schemes, because these could counteract the lack of enhancement of non-local correlations. To describe exchange-correlation, DFT-D combines the asymptotic atom-centered form with a semi-local GGA account. It also introduces fitting parameters to be used in combination with a specific GGA flavors.

%fitting parameters of DFT-D is adjusted to the specific flavor of GGA. Thus, DFT-D could implicitly provide an enhancement through its empirical fitting and the flavor of GGA. Moreover, if the underestimated $C_6$ coefficients of vdW-DF are replaced by those used in DFT-D schemes, it would reduce the difference between the atomistic pair-potential result, and the non-local correlation of vdW-DF. 

%For force field methods and DFT-D, the enhancement of non-local correlation apparently suggests that if coefficients are calculated or measured at asymptotic separations, they produce poor binding energies. Should the opposite strategy be taken and values of $C_6$ are fitted to equilibrium structures, it instead suggests too strong attractions at large separations. 

%The argument is that vdW-DF approximately includes semi-local correlation.\cite{Dion:vdW}.

\begin{figure}[h]
\centering
\includegraphics[width=8.5cm]{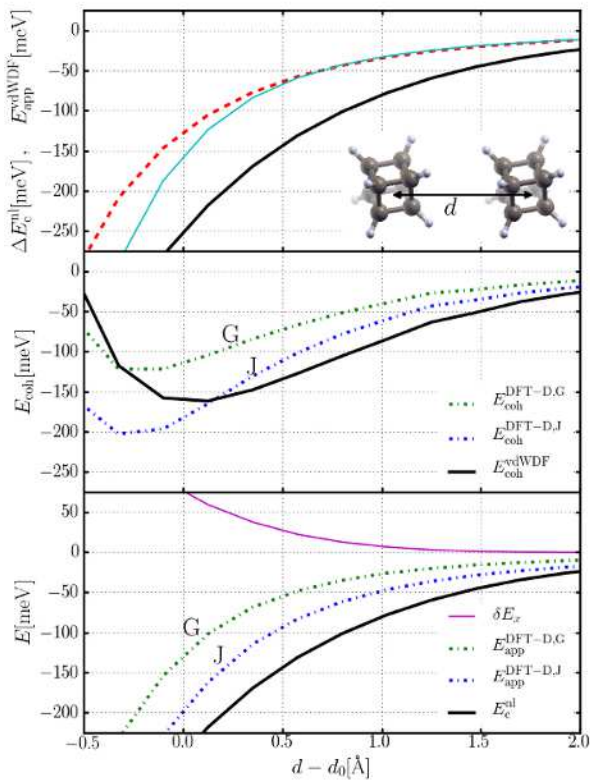}
\caption{Comparison between different accounts of non-local correlation for a dimer of cubane. In the upper panel, the dashed curve gives APP-vdWDF. The thin solid curve gives the sum of APP-vdWDF and the correlation of PBE. The thick solid curve gives the non-local correlation of vdW-DF. In the middle panel, the thick curve gives the cohesion energy using vdW-DF, $E^{\rm vdWDF}_{\rm coh}(d)$, the upper (lower) dash-dotted curve gives the DFT-D binding curve as given by Grimme, $E^{\rm DFT-D,G}_{\rm coh}(d)$ (Jure\v{c}ka, $E^{\rm DFT-D,J}_{\rm coh}(d)$). The lower panel gives the corresponding non-local correlation of vdW-DF and Grimme (Jure\v{c}ka). The thin solid curve gives the difference between exchange of revPBE and PBE. The horizontal axis gives the difference from the vdW-DF crystal binding separation $d_0$ for cubane. 
} 
\label{fig:C62}
\end{figure}

%and compare this result with the non-local correlation of vdW-DF. 
The top panel of Fig.~\ref{fig:C62} shows that gradient corrections does not account for the difference between the full vdW-DF and the APP-vdWDF results. For APP-vdWDF, we can combine the purely semi-local correlation of PBE with APP-vdWDF (in a new description APP-mod),
\begin{equation}
E^{\rm vdWDF}_{\rm APP-mod}(d) = E^{\rm vdWDF}_{\rm APP}(d)  + \Delta E^{\rm PBE}_{c}(d) - \Delta E^{\rm LDA}_{\rm c}(d)\,,
\end{equation}
to assess the magnitude of purely semi-local corrections relative to the difference between vdW-DF and APP-vdWDF. In this APP account, we have introduced the LDA and PBE terms: $\Delta E_{\rm c} (d) =E_{\rm c} (d)-E_{\rm c} (d\rightarrow \infty)$. 

We focus our discussion on the cubane dimer. The lower thin solid curve gives APP-mod; the thick solid curve gives the non-local correlation of vdW-DF. The curves are shown as a function of the intermolecular distance $d$, with $d_0$ indicating the vdW-DF binding distance in the crystal. At short separations, the thin curve lies closer to the thick curve than the corresponding APP-vdWDF result (thick dashed curve). Thus, some of the difference between APP-vdWDF and the non-local correlation arises from a lack of semi-local correlation contributions in APP-vdWDF.\cite{Dion:vdW} However, even with this inclusion the difference is still significant, and at intermediate separations it remains undiminished. 
%Note that in practical implementation of DFT-D, the long-range interactions are also damped. This damping would somewhat increase the difference between APP-vdWDF and the non-local correlation of vdW-DF. 

The middle panel of Fig. \ref{fig:C62} details the effects of using the semi-empirical fitting of DFT-D and the larger $C_6$ coefficients. We compare the binding curve for cubane obtained by use of vdW-DF with the binding curve obtained with DFT-D calculations. We select the schemes of Grimme\cite{emp:Grimme} and Jure\v{c}ka,\cite{emp:Petr} as these provide parameters for the PBE flavor of exchange-correlation, which is available to us. \cite{comment2} 

The DFT-D scheme of Grimme scales the strength of the dispersive interaction to a particular semi-local exchange-correlation to achieve good performance at short separations. For the PBE flavor of GGA, the cohesion energy of a dimer reads
\begin{equation}
E^{\rm DFT-D,G}_{\rm coh}=E^{\rm PBE}_{\rm coh}+s^{\rm PBE}_6\sum_{ij} f^{ij}_{\rm PBE,G}(|\br_i-\br_j|) \frac{C_6^{ij}}{|\br_i-\br_j|^6}\,,
\end{equation}
where $s_6=0.7$. This scheme therefore sacrifices the asymptotic description in favor of the a good description of binding separations. The scheme of Jure\v{c}ka instead adjust parameters of the damping function to the flavor of semi-local exchange-correlation,
\begin{equation}
E^{\rm DFT-D,J}_{\rm coh}=E^{\rm PBE}_{\rm coh}+\sum_{ij} f^{ij}_{\rm PBE,J}(|\br_i-\br_j|) \frac{C_6^{ij}}{|\br_i-\br_j|^6}\,,
\end{equation}
and ensure that $f^{ij}_{\rm PBE,J} \rightarrow 1$ for large separations. The DFT-D scheme can therefore, in principle, describe both asymptotic and binding separations. However, the question remains on how it performs for systems where characteristic separations lies between these two limits.

The binding curves of vdW-DF(revPBE) (the solid curve) and for the DFT-D scheme of Grimme(PBE) (upper dash-dotted curve) and Jure\v{c}ka(PBE) (lower dash-dotted curve) indicates an underestimation of {DFT-D} at intermediate separations.  
%The parameter $d_0$ corresponds to the lattice constant of cubane as determined with vdW-DF, and 
The minimum at negative $d-d_0$ shows that use of DFT-D would improve lattice constants over vdW-DF.\cite{KlimesComment} The two DFT-D schemes give quite differing binding energies. Both results show that a binding energy at the same magnitude as vdW-DF can be achieved, even with an atom-based asymptotic form of the attractive potential. The energy of vdW-DF is significantly larger at intermediate separations than that of Jure\v{c}ka despite that $C_6$ coefficients of asymptotic vdW-DF are underestimated (while those of Jure\v{c}ka are likely to be somewhat overestimated) and despite that the corresponding DFT-D scheme binds stronger than vdW-DF.

The lower panel of  Fig. \ref{fig:C62} shows that the difference at intermediate separations comes primarily from the varying accounts of correlation. The solid curve shows the non-local correlation of vdW-DF.  The dash-dotted upper (lower) curve shows
\begin{align}
E_{\rm app}^{\rm DFT-D,G(J)}&=\sum_{ij} f^{ij}_{\rm PBE,G(J)}(|\br_i-\br_j|)\frac{C_6^{ij}}{|\br_i-\br_j|^6} \nonumber \\&+\Delta E^{\rm PBE}_c(d)-\Delta E^{\rm LDA}_c(d)\,,
\end{align} 
which for DFT-D corresponds best to the non-local correlations provided by $\Delta E^{\rm nl}_{\rm c}$ in vdW-DF. The figure also shows that the difference partly cancels, at short but not at intermediate separations, with the energy difference between the exchange flavors of revPBE and PBE, $\delta E_{\rm x}= \Delta E^{\rm revPBE}_{\rm x}-E^{\rm PBE}_{\rm x} $. Thus, for certain exchange functionals, adding an asymptotic atom-based account of non-local correlation can generate good binding values, yet our vdW-DF results indicate that this framework is not optimal for describing interactions at intermediate separations.

 Our results suggest that APPs and DFT-Ds could be improved at intermediate separations. A possible strategy is to replace the atomic separations $r$ by an effective separation $r-r_0$, where $r_0$ reflects the image-planes found for nanotubes and surfaces in  Refs.~\onlinecite{ZarKohn,C6:Kleis}. Keeping this (surface-physics) effect would increase the strength of the dispersion interactions at shorter separations. 

{\it In summary}, the results of this section suggests that an asymptotic atom-based pair potentials has a limited transferability over the full range of separations. Thus, for schemes using such a form, our results raises questions on their ability to generate accurate results under broad condition (having multiple characteristic separations), for instance involving phase transitions or processes that drive the system out off equilibrium, in protein unfolding, in phase transitions, or simply for systems which have competing interactions.\cite{Lehn:Supra}  We note that there is no guarantee that vdW-DF, in its current form, can provide an accurate account under such broad conditions. A vdW-DF limitation is here  exemplified by the likely underestimation of $C_6$ coefficients for the platonic molecules. Nevertheless, we argue that non-local functionals, like vdW-DF, hold the most promise for dealing with molecular configurations under broad conditions. This is because an electron-based approach provides a framework which naturally includes image-plane and multipole effects. It therefore holds the key to an account which describe the variation in dispersive response over the full range of separations.
 
\section{Conclusions}
For the three cage-like organic molecular crystals, hexamine, cubane and dodecahedrane, vdW-DF gives lattice parameters and cohesion energy that agree well with experiments, although for all three crystals the unit cell volumes are overestimated. A substantial sensitivity of lattice parameters and cohesion energy to the flavor of semi-local exchange signals that this overestimation stems mostly from the chosen form of exchange functional. 

We have also shown that, at short and intermediate separations, the full non-local correlation of vdW-DF is considerably larger than its corresponding atom-based asymptotic account. Notwithstanding that the asymptotic account of vdW-DF likely needs improvement, this enhancement indicates that the asymptotic $1/r^6$ form of atomic pair potentials, by construction, can not give a transferable account over a large range of separations. 

This paper underlines the usefulness of studying simple model sparse systems as molecular crystals to gain insight into methods intended for the study of sparse and supramolecular systems.\cite{Lehn:Supra} Both DFT-D and vdW-DF benefit from such testing, because they are designed to be parameter-free, and an accurate account of molecular crystals would suggest an accurate account of more complex assemblies of similar molecules. 
The presented molecular crystals provide particularly accessible cases of extended systems and can therefore be used in conjunction with future development of pair-potential methods and exchange-correlation functionals. 

\section{Acknowledgements}
We acknowledge E. Schr\"oder for useful comments and discussions. We thank I. Sinno for help in designing the molecular-crystal schematics and {\O}.\ Borck for access to his code for Bader analysis. The Swedish National Infrastructure for Computing (SNIC) is acknowledged for computer allocation and for  KB’s  participation in the graduate school NGSSC. The work was supported by the  Swedish research  Council (Vetenskapsr\aa det VR) under  621-2008-4346.


\begin{thebibliography}{5}
\bibitem{comment3} We have explicitly verified that the precise assignment of the total molecular response onto the atoms does not alter the APP-vdWDF curves significantly. An assignment of the total asymptotic vdW-DF response according to the relative strength of Grimme yield a weak increase in the difference between APP-vdWDF and the non-local correlation.
\bibitem{comment2} In our comparison, we neglect that these schemes are optimized for a particular basis set, arguing that our comparison does not require a fine-tuned version of DFT-D.
\end{thebibliography}
\end{document}